\documentclass[12pt]{article}
\usepackage{amsfonts}
\usepackage{amssymb}
\usepackage{amsmath}
\usepackage{graphics}
\begin{document}
{\LARGE \centerline{Quantum canonical transformations}
\centerline{in}\centerline{Weyl-Wigner-Groenewold-Moyal
formalism}}

\phantom{aaa}

\phantom{aaa}

{\large \centerline{T. Dereli}}%
\centerline{Physics Department, Ko\c{c} University, 80910
Sar{\i}yer-Istanbul, TURKEY}%
\centerline{{\it tdereli@ku.edu.tr}}

{\large \centerline{T. Hakio\u{g}lu}}%
\centerline{Physics Department, Bilkent University, 06533 Ankara, TURKEY}%
\centerline{{\it hakioglu@fen.bilkent.edu.tr}}

{\large \centerline{A. Te\u{g}men\footnote{On sabbatical from
Physics Department,
Ankara University 06100 Ankara TURKEY}}}%
\centerline{Feza G\"{u}rsey Institute, 34684
\c{C}engelk\"{o}y-Istanbul,
TURKEY}%
\centerline{{\it tegmen@science.ankara.edu.tr}}

\begin{abstract}
A conjecture in quantum mechanics states that any quantum
canonical transformation can decompose into a sequence of three
basic canonical transformations; gauge, point and interchange of
coordinates and momenta. It is shown that if one attempts to
construct the three basic transformations in star-product form,
while gauge and point transformations are immediate in
star-exponential form, interchange has no correspondent, but it is
possible in an ordinary exponential form. As an alternative
approach, it is shown that all three basic transformations can be
constructed in the ordinary exponential form and that in some
cases this approach provides more useful tools than the
star-exponential form in finding the generating function for given
canonical transformation or vice versa. It is also shown that
transforms of $c$-number phase space functions under
linear-nonlinear canonical transformations and intertwining method
can be treated within this argument.
\end{abstract}
\noindent\rule{7in}{0.01in}
\section{Introduction}
Weyl-Wigner-Groenewold-Moyal (WWGM) formalism provides us a
quantization and dequantization scheme based on Weyl's
correspondence \cite{ref:Weyl} and Wigner's quasi-distribution
function \cite{ref:Wigner} between quantum mechanical operators
and $c$-number phase space functions. The product rule of these
functions is given by Groenewold-Moyal's so-called twisted or star
product ($\star$-product) \cite{ref:Groenewold,ref:Moyal}. For a
comprehensive treatment of the subject the reader may consult
Refs. \cite{ref:Schroek}-\cite{ref:ZFC}.

Since the playground of quantum canonical transformations (QCTs)
is the quantum phase-space it is natural to introduce connection
between QCTs and their $c$-number phase-space picture. Starting
with the pioneering works of B. Leaf \cite{ref:Leaf}, various
aspects of this subject have been studied in the literature much
considering behavior of the Wigner function under CTs. An
extensive list of references can be found in Ref. \cite{ref:ZFC}.

Surprisingly, two independent fundamental types of invertible
phase space maps in one variable were proposed as the elementary
generators of the entire classical and QCTs, that is every CT can
be decomposed as finite or infinite sequences of the elementary
CTs \cite{ref:Leyvraz}. These are linear and point CTs. Later
elaborations of this conjecture in quantum mechanics led to a
triplet as a wider class including gauge transformations, point
transformations and finally interchange of coordinates and momenta
\cite{ref:Deenen,ref:Anderson}. As a crash problem, this statement
has not been proven in a general framework yet. But, though should
it is not true for every CT it applies to a large and relevant
class of CTs.

The present work deals mainly with the implementation of the
conjecture stated above in WWGM formalism. First we give a brief
summary on the fundamental QCTs following the Refs.
\cite{ref:Deenen},\cite{ref:Anderson}. In Sec. 2, we see that if
we use the algebra isomorphism between the Hilbert space operators
and the $c$-number phase-space functions, gauge and point
transformations appear immediately in $\star\,$- exponential form
just as in the expected form, but the interchanging remains out of
this isomorphism. Still, we will be able to construct the
interchanging in an ordinary exponential form. By accepting this
result as our main guide, Sec. 3 is devoted to show that other two
fundamental CTs can also be set on an isomorphism independent
background. Sec. 4 shows that the generators in the ordinary
exponential form are compatible with the well-known behaviors of
functions under both linear and non-linear CTs. In Sec. 5, after
construction of the intertwining method in terms of
$\star$-product we emphasize that the intertwining equation may be
used to determine the relation between non-intertwined potentials
well. Finally Sec. 6 contains a short summary and conclusions.

A QCT is defined as
\begin{eqnarray}\label{QCT}
\hat{F}(\hat{q},\hat{p})\;\hat{q}\;\hat{F}^{-1}(\hat{q},\hat{p})=
\hat{Q}(\hat{q},\hat{p}), \quad
\hat{F}(\hat{q},\hat{p})\;\hat{p}\;\hat{F}^{-1}(\hat{q},\hat{p})=
\hat{P}(\hat{q},\hat{p}),
\end{eqnarray}
where $[\,\hat{Q},\hat{P}\,]=\hat{Q}\hat{P}-\hat{P}\hat{Q}=i\hbar$
and $\hat{F}(\hat{q},\hat{p})$ is the generating function (GF)
which is an arbitrary complex function (unitary or non-unitary),
and $\hat{F}^{-1}$ is the algebraic inverse of $\hat{F}$. Action
of the exponential version of the transformation on an arbitrary
quantum phase-space function $\hat{u}(\hat{q},\hat{p})$ is given
by the well-known form
\begin{eqnarray}\label{expquantum}
e^{\lambda \hat{f}}\;\hat{u}(\hat{q},\hat{p})\;e^{-\lambda
\hat{f}}=\hat{u}+\lambda
[\;\hat{f},\hat{u}\;]+\frac{\lambda^2}{2!}[\;\hat{f},[\;\hat{f},\hat{u}\;]\;]+
\frac{\lambda^3}{3!}[\;\hat{f},[\;\hat{f},[\;\hat{f},\hat{u}\;]\;]\;]+\cdots
,
\end{eqnarray}
where $\lambda$ is a pure imaginary number with a continuous
parameter. The gauge transformation is generated, via
(\ref{expquantum}), by complex function $f(\hat{q})$:
\begin{eqnarray}\label{gt}
e^{\lambda f(\hat{q})}\;\hat{q}\;e^{-\lambda f(\hat{q})}= \hat{q},
\quad e^{\lambda f(\hat{q})}\;\hat{p}\;e^{-\lambda
f(\hat{q})}=\hat{p}+i\hbar \lambda\;
\partial_q f,
\end{eqnarray}
where $\partial_q\equiv \frac{\partial}{\partial q}$.

The point CT (change of variables) is given by
\begin{eqnarray}\label{pt}
e^{\lambda f(\hat{q})\,\hat{p}}\;\hat{q}\;e^{-\lambda
f(\hat{q})\,\hat{p}}= A(\hat{q}), \quad e^{\lambda
f(\hat{q})\,\hat{p}}\;\hat{p}\;e^{-\lambda
f(\hat{q})\,\hat{p}}=(\partial_q A)^{-1}\,\hat{p},
\end{eqnarray}
where
\begin{eqnarray}\label{A}
A(\hat{q})=e^{-i\hbar\lambda f(\hat{q})\partial_q}\;\hat{q} .
\end{eqnarray}
In (\ref{pt}), it is immediate to see that when the order in
$f(\hat{q})\,\hat{p}$ is reversed, the order in $(\partial_q
A)^{-1}\,\hat{p}$ is reversed.

Finally, the interchange of coordinates and momenta
\begin{eqnarray}\label{def-int}
\hat{I}\,\hat{q}\,\hat{I}^{-1}=\hat{p},\quad
\hat{I}\,\hat{p}\,\hat{I}^{-1}=-\hat{q}
\end{eqnarray}
is achieved by the Fourier transform operator $\hat{I}$ whose
definition is given by the action
\begin{eqnarray}\label{action}
\hat{I}^{\pm
1}\,f(\hat{q})=\frac{1}{(2\pi)^{1/2}}\int_{-\infty}^{\infty}f(\hat{q}')\,e^{\pm
i\hat{q}\hat{q}'/\hbar}\;dq'.
\end{eqnarray}
As a special case of the linear CT, the interchanging can also be
constructed by the composition of gauge transformations
\cite{ref:Anderson}
\begin{eqnarray}\label{inter}
\hat{F}_I(\hat{q},\hat{p})=e^{i\hat{q}^2/(2\hbar)}\;
e^{i\hat{p}^2/(2\hbar)}\;e^{i\hat{q}^2/(2\hbar)}.
\end{eqnarray}
But because of the linearity property,
\begin{eqnarray}
\hat{I}\hat{u}(\hat{q},\hat{p})\,\hat{I}^{-1}=\hat{u}(\hat{p},-\hat{q}),
\end{eqnarray}
the middle term in (\ref{inter}) can not be accepted as an
independent transformation within the class composed of gauge,
point and interchange transformations. But note that, at the
beginning it is possible to give the gauge (\ref{gt}) and point
(\ref{pt}) transformations so as to be based on function
$f(\hat{p})$ such as $e^{\lambda f(\hat{p})}$ and $e^{\lambda
f(\hat{p})\,\hat{q}}$ respectively. Throughout the text our choice
will be $f(\hat{q})$.

On the other hand, the linear CT itself can be decomposed into the
form \cite{ref:Anderson}
\begin{eqnarray}\label{LT}
\hat{F}_L(\hat{q},\hat{p})&=&e^{\gamma \hat{q}\hat{p}}\;
e^{\beta \hat{q}^2}\;e^{\alpha \hat{p}^2}\nonumber\\
&=&e^{\gamma \hat{q}\hat{p}}\; e^{\beta
\hat{q}^2}\;\hat{I}\,e^{\alpha \hat{q}^2}\,\hat{I}^{-1},
\end{eqnarray}
where $\alpha , \beta , \gamma $ are the pure imaginary numbers
compatible with a linear CT, therefore linear CT is not an element
of the class defined above.
\section{Implementations in WWGM Formalism}
Given a $c$-number phase-space monomial $q^m\,p^n$ with
non-negative integers $m$, $n$, its image in the Hilbert space as
a symmetrically ordered operator is determined by the Fourier
transform
\begin{equation}\label{WC}
\hat{F}(\hat{q},\hat{p})=
\frac{1}{(2\pi)^2}\int_{-\infty}^{\infty}d\sigma\, d\tau\, dq\,
dp\, F(q,p)\;e^{i\,[\sigma (\hat{q}-q)+\tau (\hat{p}-p)]/\hbar}
\end{equation}
which serves as the quantization procedure called Weyl
quantization \cite{ref:Weyl}. Conversely, given operator
$\hat{F}(\hat{q},\hat{p})$, the phase-space kernels $F(q,p)$ are
specified simply by the correspondence
\begin{eqnarray}
\hat{q}\rightarrow q,\quad \hat{p}\rightarrow p
\end{eqnarray}
provided that $\hat{F}(\hat{q},\hat{p})$ is symmetrically ordered.

The associative (but non-abelian in general) $\star$-product
corresponding to the operator product in the Hilbert space is
given by
\begin{eqnarray}
\star\;=\,e^{\frac{i\,\hbar}{2}(\overleftarrow{\partial}_q\overrightarrow{\partial}_p-
\overleftarrow{\partial}_p\overrightarrow{\partial}_q)},
\end{eqnarray}
where the arrows indicate the direction that the derivatives act.
$\star$-product of $c$-number phase-space monomials and operator
product of their images are in a complete algebra isomorphism
given by the equation \cite{ref:Groenewold}
\begin{eqnarray}
\hat{F}(\hat{q},\hat{p})\;\hat{G}(\hat{q},\hat{p})=
\frac{1}{(2\pi)^2}\int_{-\infty}^{\infty}d\sigma\, d\tau\, dq\,
dp\, [F(q,p)\star G(q,p)]\;e^{i\,[\sigma (\hat{q}-q)+\tau
(\hat{p}-p)]/\hbar}.
\end{eqnarray}
For example, while $q\star p^2$ is going to $\hat{q}\,\hat{p}^2$,
$qp\star p$ goes to $\hat{q}\hat{p}^2-i\hbar \hat{p}/2$ which is
equivalent to the product of symmetrically ordered images of $qp$
and $p$. By means of this isomorphism, it is possible to make
practically some simple (de)- quantization operations. For
example, the operator $\hat{p}\,\hat{q}$ which is not
symmetrically ordered is the quantized version of $p\star
q=qp-i\hbar/2$. Conversely, quantized version of the $c$-number
function $qp$=$(q\star p+p\star q)/2$ is
$(\hat{q}\hat{p}+\hat{p}\,\hat{q})/2$. Therefore these examples
induce that the Weyl quantization procedure of a $c$-number
function is automatically reduced to write it in terms of the
$\star$-product, meanwhile the de-quantization procedure is easier
obviously. Although this (de)-quantization scheme as a unitary
mapping is restricted to the monomials, it can be set for more
general classes of functions and operators
\cite{ref:Dubin,ref:ZFC}. Now it has been recently understood that
this correspondence can be achieved in terms of a kernel function
\cite{ref:Strat,ref:Brif}. In these general terms, for the sake of
generality, we assume that there always exists a one to one
correspondence between arbitrary $F(q,p)$ and
$\hat{F}(\hat{q},\hat{p})$. If this general correspondence which
can be summarized as
\begin{eqnarray}\label{correspondence}
F(q,p)&\leftrightarrow  &\hat{F}(\hat{q},\hat{p}),\nonumber\\
F(q,p)\star G(q,p) &\leftrightarrow &\hat{F}(\hat{q},\hat{p})
\,\hat{G}(\hat{q},\hat{p})
\end{eqnarray}
is used for the QCT (\ref{QCT}), the corresponding transformation
in the $c$-number phase-space can be written as: \cite{ref:TH}
\begin{eqnarray}\label{defn}
F(q,p)\star q\star F^{-1}(q,p)=Q(q,p),\qquad F (q,p)\star p\star
F^{-1}(q,p)=P(q,p)
\end{eqnarray}
satisfying
\begin{eqnarray}\label{PMcond}
\{\,Q\,,P\,\}^M=i\hbar
\end{eqnarray}
and $F \star F ^{-1}=F^{-1}\star F=1$, where
$\{\,Q\,,P\,\}^M=Q\star P-P\star Q$ is the Moyal bracket and
$F^{-1}(q,p)$ is the algebraic inverse of the GF $F(q,p)$.

If we employ the facts that $\{\,F\,,q\,\}^M=i\hbar\,\partial_p F$
and $\{\,F\,,p\,\}^M=-i\hbar\,\partial_q F$, we can write the
definition (\ref{defn}) in a more useful form
\begin{subequations}\label{defnalt}
\begin{equation}
Q(q,p)=q-i\hbar\,\partial_p F(q,p)\star F^{-1}(q,p),
\end{equation}
\begin{equation}
\quad P(q,p)=p+i\hbar\,\partial_q F(q,p)\star F^{-1}(q,p).
\end{equation}
\end{subequations}
The correspondence (\ref{correspondence}) implies the gauge
transformation (\ref{gt}) in WWGM formalism as
\begin{eqnarray}
e_\star^{\lambda f(q)}\star q\star e_\star^{-\lambda f(q)}= q,
\quad e_\star ^{\lambda f(q)}\star p\star e_\star^{-\lambda
f(q)}=p+i\hbar \lambda\;
\partial_q f,
\end{eqnarray}
which gives the de-quantized form of (\ref{gt}), where the
$\star$-exponential is given by \cite{ref:Imre}-\cite{ref:Bayen}
\begin{eqnarray}\label{starexp}
e_\star^{\lambda\,f(q,p)}=1+\lambda\,f(q,p)+\frac{\lambda^2}{2!}\,f(q,p)\star
f(q,p)+\cdots\;.
\end{eqnarray}
When the property
\begin{eqnarray}
\partial_\lambda\,e_{\star}^{\lambda\,f(q,p)}=f(q,p)\star e_{\star}^{\lambda\,f(q,p)}
=e_{\star}^{\lambda\,f(q,p)}\star f(q,p).
\end{eqnarray}
is performed, one can obtain
\begin{eqnarray}\label{expstar}
e_\star^{\lambda f(q,p)}\star \;u(q,p)\star \;e_\star^{-\lambda
f(q,p)}&=&\sum_{n,r=0}^ \infty\left(
\begin{array}{c}
n\\
r \end{array} \right) (-1)^r\frac{\lambda^{n+r}}{(n+r)!}\,
(f\star )^{\,n}\,u\,(\star f)^{\,r}\nonumber\\
&=&u+\lambda\;
\{\;f,u\;\}^M+\frac{\lambda^2}{2!}\{\;f,\{\;f,u\;\}^M\;\}^M\nonumber\\
&+&\frac{\lambda^3}{3!}\{\;f,\{\;f,\{\;f,u\;\}^M\;\}^M\;\}^M+\cdots\,.
\end{eqnarray}
Then the point CT (\ref{pt}) amounts to
\begin{eqnarray}\label{ptm}
e_\star^{\lambda f(q)\star p}\star q\star e_\star^{-\lambda
f(q)\star p}= A(q), \quad e_\star^{\lambda f(q)\star p}\star
p\star e_\star^{-\lambda f(q)\star p}=(\partial_q A)^{-1}\star p,
\end{eqnarray}
where $A(q)$ is as in (\ref{A}). In attempting to construct the
interchange GF (\ref{action}) it should be considered the fact
that the $c$-number phase-space GFs are not operators. Therefore
searching a function taking integral (or maybe taking derivative)
is meaningless and such a function is not available in the
$\star$-product argument within the isomorphism given above.
Still, things can be put right by converting (\ref{defn}) into the
system of partial differential equations \cite{ref:THmain}
\begin{eqnarray}\label{interchange}
F_I(q,p)\star\,q=p\star\, F_I(q,p),\qquad
F_I(q,p)\star\,p=-q\star\, F_I(q,p).
\end{eqnarray}
The solution
\begin{eqnarray}\label{im}
F_I(q,p)=e^{i(q^2+p^2)/\hbar}
\end{eqnarray}
can be accepted as the GF searched for the last member of the set
of fundamental transformations. Note that $F_I(q,p)$ appears in an
ordinary exponential form not in the $\star$-exponential form like
the others. Therefore it may be natural to ask whether each of the
three basic GFs can be obtained in the ordinary exponential form.
The proceeding section is devoted to discuss this approach.

\section{Ordinary Exponential versus $\star$-exponential}
(\ref{defn}) is used for two main purposes; given GF, finding the
CT, i.e., the pair $(Q(q,p),P(q,p))$ and vice versa. When the
transformations are defined in terms of the $\star$-exponential
form, given GF determination of $Q(q,p)$ and $P(q,p)$ is much
easier, since the expansion (\ref{expstar}) is a powerful tool.
But conversely, given a CT, determination of the GF is generally a
tedious matter. On the other hand, it is always possible to
convert (\ref{defn}) into a system of partial differential
equations for the GFs in the ordinary exponential form;
\begin{eqnarray}\label{tech}
e^{\lambda f(q,p)}\star q=Q(q,p)\star e^{\lambda f(q,p)},\quad
e^{\lambda f(q,p)}\star p=P(q,p)\star e^{\lambda f(q,p)},
\end{eqnarray}
which is a modified definition of the CTs that can be used to find
both the GF and the pair $(Q(q,p),P(q,p))$. As a remarkable point
note that, in operating (\ref{tech}) to find the CT, one does not
need to know the inverse of the GF.

Alternatively, if the CT is defined simply by eliminating the star
sign, (\ref{expstar}) is deformed to strictly different expansion
\begin{eqnarray}\label{expnormal}
e^{\lambda f(q,p)}\star u(q,p)\star e^{-\lambda
f(q,p)}=\sum_{n,r=0}^ \infty\left(
\begin{array}{c}
n\\
r \end{array} \right) (-1)^r\frac{\lambda^{n+r}}{(n+r)!}
f^{\,n}\star u\star f^{\,r}.
\end{eqnarray}
This form behaves in a contrast way such that the right hand side
of (\ref{expnormal}) is not so easy to evaluate. In order to get
rid of this problem, one may attempt to convert (\ref{expnormal})
into (\ref{tech}). This is always possible when $f(q,p)=f(q)$ or
$f(q,p)=f(p)$ because with this condition the term $exp[-\lambda
f(q,p)]$ in (\ref{expnormal}) is always the inverse of
$exp[\lambda f(q,p)]$. Thus the definitions (\ref{tech}) and
(\ref{expnormal}) becomes equivalent. Since the equations
\begin{eqnarray}\label{remark}
e_\star^{\lambda\,f}=e^{\lambda\,f},\quad (f\star
)^{\,n}=f^{\,n}\star
\end{eqnarray}
also hold, $e_\star^{\lambda\,f}$ and $e^{\lambda\,f}$ generate
the same CT via (\ref{expstar}) or (\ref{expnormal}) equivalently.

These remarks allow us to use the ordinary exponential form to
generate the gauge transformation directly:
\begin{eqnarray}\label{gauge}
e^{\lambda f(q)}\star q\star e^{-\lambda f(q)}= q, \quad e
^{\lambda f(q)}\star p\star e^{-\lambda f(q)}=p+i\hbar \lambda\;
\partial_q f.
\end{eqnarray}
Conversely, given gauge transformation $Q=q$, $P=p+u(q)$, the GF
\begin{eqnarray}
F_G(q,p)=e^{-\frac{i}{\hbar}\int u(q)\,dq}
\end{eqnarray}
appears as the solution to system of partial differential
equations
\begin{eqnarray}
-\partial_p\, F_G=\partial_p\, F_G,\quad i\hbar\, \partial_q
F_G=u( q+i\hbar\partial_p /2) F_G
\end{eqnarray}
which is obtained from (\ref{tech}).

On the other hand, according to (\ref{defn}) and the canonicity
condition (\ref{PMcond}), the most general form of the point
transformation must satisfy the system of partial differential
equations
\begin{subequations}\label{ord-point}
\begin{equation}\label{ord-point1}
F_P(q,p)\star q = Q(q)\star F_P(q,p),
\end{equation}
\begin{equation}\label{ord-point2}
F_P(q,p)\star p = [\,\tilde{Q}(q)\,p+\chi (q)\,]\star F_P(q,p),
\end{equation}
\end{subequations}
where $\tilde{Q}(q)=[\,\partial_q Q(q)\,]^{-1}$ and $\chi (q)$,
for the time being, is an arbitrary function. The system
(\ref{ord-point}) may be solved by looking for solutions of the
form
\begin{eqnarray}
F_P(q,p)=e^{\lambda [\,f(q)p+g(q)\,]}.
\end{eqnarray}
Indeed, consider the facts that
\begin{eqnarray}
F(q)\star e^{G(q)p+H(q)}=F[\,q+i\hbar\,G(q)/2\,]\,e^{G(q)p+H(q)},
\end{eqnarray}
\begin{eqnarray}
\frac{\partial}{\partial
q}F[\,\underline{q}+G(q)\,]=\frac{\partial G}{\partial
q}\frac{\partial F}{\partial q}[\,q+G(q)\,],
\end{eqnarray}
where $F, G, H$ are arbitrary functions and $\underline{q}$ is
considered as constant under the operation $\partial_q$, which is
originated from a crucial property of the $\star$-product. Then
(\ref{ord-point1}) requires
\begin{eqnarray}\label{Q}
Q(\upsilon)=q-\frac{i\hbar\lambda}{2}\,f,
\end{eqnarray}
where $\upsilon =q+i\hbar\lambda f/2$. The equality of the
coefficients of $p\,$s with equal powers on both sides of
(\ref{ord-point2}) requires
\begin{eqnarray}
\tilde{Q}(\upsilon)=\left[\frac{\partial Q}{\partial
q}(\upsilon)\right]^{-1}= \frac{2+i\hbar\lambda\partial_q
f}{2-i\hbar\lambda\partial_q f}\,,
\end{eqnarray}
\begin{eqnarray}\label{chi}
\chi
(\upsilon)=\frac{i\hbar\lambda}{2}\left[1+\tilde{Q}(\upsilon)\right]\frac{\partial
g}{\partial q}-\frac{\lambda\,\hbar^2}{4}\frac{\partial
\tilde{Q}}{\partial q}(\upsilon)\frac{\partial f}{\partial q}.
\end{eqnarray}
Given CT, i.e., $Q(q)$ and $\chi (q)$ or GF, i.e., $f(q)$ and
$g(q)$, (\ref{Q}) and (\ref{chi}) provide analytical or numeric
solutions. We now examine some important cases. \newline\newline%
(i) $f=c_1$ and $g=g(q)$, where $c_1$ is any constant. Such a
choice gives
\begin{eqnarray}
Q(q)=q-i\hbar\lambda\,c_1,
\end{eqnarray}
\begin{eqnarray}
P(q,p)=p+i\hbar \lambda\,\frac{\partial g}{\partial
q}(q-i\hbar\lambda\,c_1/2)
\end{eqnarray}
due to (\ref{Q}) and (\ref{chi}) respectively. If $c_1=0$, then
this is the gauge transformation (\ref{gauge}). This result shows
obviously that the gauge transformation is a special case of the
point transformation and it makes the gauge transformation
unnecessary as an independent fundamental transformation so long
as the point transformation is defined by (\ref{ord-point}) with
$\chi (q)\neq 0$. On the other hand, one may define the point
transformation so as to be $\chi (q)=0$ without destroying the
canonicity condition (\ref{PMcond}). With this definition, the
gauge transformation becomes a necessary member of the class of
fundamental transformations. Now, (\ref{Q}) is still valid and
(\ref{chi}) gives
\begin{eqnarray}\label{g}
g(q)=-\frac{i\hbar}{2}\int \frac{\frac{\partial
\tilde{Q}}{\partial
q}(\upsilon)}{1+\tilde{Q}(\upsilon)}\,\frac{\partial f}{\partial
q}\,dq.
\end{eqnarray}
If $f(q)$ and also $g(q)$ are chosen arbitrarily as nonconstant
functions, according to (\ref{chi}) we see that the existence of
$\chi (q)$ in $P(q,p)$ becomes generally inevitable. The following
example may make these points more clear.

Consider the CT $Q=1/q$, $P=-q^2 p$. (\ref{Q}) and (\ref{chi}) (or
(\ref{g})) give rise to
\begin{eqnarray}\label{f}
f(q)=\pm\frac{2}{\hbar\lambda}(1-q^2)^{1/2}
\end{eqnarray}
and
\begin{eqnarray}
g(q)=-\frac{1}{2\lambda}\ln (q^2-1)
\end{eqnarray}
respectively, where $g(q)$ is evaluated for the $f(q)$ with
positive sign. Conversely, given GF containing the same $f(q)$
with the positive sign in (\ref{f}) and $g=0$, we get the CT as
\begin{eqnarray}
Q(q)=1/q,\quad P(q,p)=-q^2p-\frac{i\hbar}{q}\frac{1+q^2}{1-q^2}.
\end{eqnarray}
(ii) $f=f(q)$ and $g=c_2$, where $c_2$ is any constant.

(a) $f=q$. (\ref{Q}) and (\ref{chi}) amount to the scaling
transformation
\begin{eqnarray}\label{st}
Q(q)=k\,q,\quad P(q,p)=\frac{1}{k}\,p
\end{eqnarray}
with $\chi =0$, where $k=(2-i\hbar\lambda )/(2+i\hbar\lambda )$,
($\lambda \neq 2i/\hbar )$. Note that the scaling transformation
(\ref{st}) is compatible with (\ref{ptm}) for a different $A(q)$.
Since
\begin{eqnarray}
e_\star^{\mu q\star p}\star q\star e_\star^{-\mu q\star p}=
e^{-i\hbar\mu}q, \quad e_\star^{\mu q \star p}\star p\star
e_\star^{-\mu q\star p}=e^{i\hbar\mu}\star p=e^{i\hbar\mu}p,
\end{eqnarray}
where $\mu$ is a pure imaginary number, (\ref{st}) can be
generated by the $\star$-exponential function
\begin{eqnarray}\label{scaling}
e_\star^{\frac{i}{\hbar}\,(\ln k)\,q\star p}
\end{eqnarray}
corresponding to the quantized form of $e^{\lambda qp}$. This
result, i.e.,  $e^{i\,(\ln k)\,\hat{q}\hat{p}/\hbar}$, is not
immediate if one attempts to quantize $e^{\lambda\, q\,p}$ using
the Weyl correspondence (\ref{WC}).

(b) $f=q^2$ generates the CT
\begin{eqnarray}
Q(q)=-q+\frac{2i}{\hbar\lambda}\left( 1 \pm
\eta \right),\nonumber\\
P(q,p)=-\frac{\eta}{(\eta + 2)}p -\lambda\,\hbar^2 \frac{\eta +1
}{\eta (\eta +2)^2},
\end{eqnarray}
where $\eta =(1+2i\hbar\lambda q)^{1/2}$ and $P(q,p)$ is evaluated
for the $Q(q)$ with positive sign.\newline\newline%
(iii)

(a) $Q(q)=\ln{q}, P(q,p)=q\,p$ which is one of the three
successive transformations in transforming the quantum Liouville
Hamiltonian to a free particle. For this, one must solve
$\;e^{q-i\hbar\lambda f(q)/2}=q+i\hbar\lambda f(q)/2\;$
numerically \cite{ref:THmain}.

(b)The inverse of (a) is $Q(q)=e^{q}, P(q,p)=e^{-q}\,p$. It is a
typical example for a spectrum non-preserving transformation that
one encounters in the phase space representations of the radial
dimension \cite{ref:THmain,ref:JOSAA}. For this transformation one
obtains $\;e^{q+i\hbar\lambda f(q)/2}=q-i\hbar\lambda f(q)/2\;$
for which a numerical solution is
necessary.\newline\newline%
(iv) Finally, one may choose
\begin{eqnarray}
g(q)=\frac{i\hbar}{2}\,\partial_q f\;,\; \qquad \chi
(q)=\frac{i\hbar}{2}\,\partial_q \tilde{Q}
\end{eqnarray}
so that the transformation becomes
\begin{eqnarray}
e^{\lambda f\star p}\star q =Q(q)\star e^{\lambda f\star p},\qquad
e^{\lambda f\star p}\star p =\tilde{Q}\star p\star e^{\lambda
f\star p}
\end{eqnarray}
which is the ordinary exponential analogous of (\ref{ptm}). But
(\ref{chi}) shows that such a transformation induces the condition
\begin{eqnarray}
\frac{\partial \tilde{Q}}{\partial q}(\upsilon)=\frac{\partial
}{\partial q}\tilde{Q}(\upsilon)
\end{eqnarray}
which is not always possible. One possible case is the scaling
transformation given above.

\section{Transform of Functions}

This section considers the behavior of phase space functions under
the linear and nonlinear CTs in turn within the ordinary
exponential form.  While the linear case can be investigated in a
general framework, the nonlinear case is given by a particular
example. We show that both results are compatible with the ones in
the literature \cite{ref:Dragt,ref:Curtright}.

The linear CTs satisfy the equation
\begin{eqnarray}
F(q,p)\star u(q,p)\star F^{-1}(q,p)=u\left(
FqF^{-1},FpF^{-1}\right) = u(Q,P),
\end{eqnarray}
for any arbitrary phase space function $u(q,p)$, which is
especially shown in the literature for the Wigner functions
\cite{ref:Dragt,ref:Zachos}. Now, we would like to show this
covariance in a general compact way keeping ourselves in the
ordinary exponential form. In doing so, we will use the Lie
operator method which is very suitable for the treatment.

The Lie operator associated with the transformation
(\ref{expstar}) is defined by \cite{ref:ADTH}
\begin{eqnarray}
\hat{L}_M&=&f\star -\star f = f( q+i\hbar\partial_p/2\;
,\,p-i\hbar\partial_q/2 ) -f( q-i\hbar\partial_p/2\;
,\,p+i\hbar\partial_q/2 ).
\end{eqnarray}
$\hat{L}_M$ acts on $u(q,p)$ such as
\begin{eqnarray}
\hat{L}_M\,u\,=\,\{f,u\}^M=f\star u-u\star f,
\end{eqnarray}
that the result is obviously a Moyal bracket. The powers of
$\hat{L}$ are given by
\begin{eqnarray}
\hat{L}_M^0\,u\,&=&\,u,\nonumber\\
\hat{L}_M\,u\,&=&\,\{f,u\}^M,\nonumber \\
\hat{L}_M^2\,u\,&=&\hat{L}_M\{f,u\}^M=\{f,\{f,u\}^M\}^M\,,
\end{eqnarray}
and so on. Thus the construction of the transformation
(\ref{expstar}) in terms of the Lie operator is straightforward,
so that
\begin{eqnarray}\label{procedure}
e_\star^{\lambda f}\star u\star e_\star^{-\lambda f}=e^{\lambda
\hat{L}_M}u.
\end{eqnarray}

The linear CT is given by
\begin{eqnarray}\label{LCT}
Q(q,p)&=&a\,q+b\,p \;,\nonumber\\
P(q,p)&=&c\,q+d\,p \; ,
\end{eqnarray}
where $a,b,c,d$ are real constants satisfying $a\,d-b\,c=1$ and
$a+d+2 \neq 0$. The compact GF as the solution of (\ref{tech}) is
given by
\begin{eqnarray}\label{LMGF}
F(q,p)=e^{2iA[\,bp^2-cq^2+(a-d)qp\,]/\hbar},
\end{eqnarray}
where $A=1/(a+d+2)$. Since the Lie operator method is based on the
$\star$-exponential, (\ref{LMGF}) is not suitable to generate the
linear CT within the procedure (\ref{procedure}). But first, let
us consider the decomposed form of the linear CT
\begin{eqnarray}\label{LCTGF}
F_L\,'(q,p)=e_\star^{\gamma\,q\star p}\star e_\star^{\beta
q^2}\star e_\star^{\alpha p^2}
\end{eqnarray}
which corresponds to (\ref{LT}). Second, if we go ahead one step
more we reach, by (\ref{remark}) and (\ref{scaling}), that
\begin{eqnarray}\label{decomp}
F_L(q,p)=e^{\lambda qp}\star e^{\beta q^2}\star e^{\alpha p^2}
\end{eqnarray}
which is a decomposition of (\ref{LMGF}) in terms of ordinary
exponentials. The uniqueness principle of the GFs allows us to use
(\ref{decomp}) instead of (\ref{LMGF}) and therefore we reach the
result
\begin{eqnarray}
F_L\star u(q,p)\star F_L^{-1}=u\left( F_L\star q\star
F_L^{-1},F_L\star p \star F_L^{-1}\right).
\end{eqnarray}
Indeed, the first movement gives
\begin{eqnarray}
e^{\alpha p^2}\star\,u(q,p)\star\,e^{-\alpha p^2}=
e^{\alpha\hat{L}_{M_1}}u(q,p)=u\left(q-2i\hbar\alpha p,p\right),
\end{eqnarray}
where $\hat{L}_{M_1}=p^2\star -\star p^2=-2i\hbar p\,\partial_q$.
The second one amounts to
\begin{eqnarray}
e^{\beta q^2}\star\,u\left(q-2i\hbar\alpha
p,p\right)\star\,e^{-\beta q^2}&=&
e^{\beta\hat{L}_{M_2}}u\left(q-2i\hbar\alpha
p,p\right)\nonumber\\
&=&u\left( (1+4\hbar^2\alpha\beta )\,q-2i\hbar\alpha p\,,
p+2i\hbar\beta \,q\right),
\end{eqnarray}
where $\hat{L}_{M_2}=q^2\star -\star q^2=2i\hbar q\partial_p$. And
finally the third one can be achieved by the Lie operator
\begin{eqnarray}
\hat{L}_{M_3}= -i\hbar q\,\partial_q +i\hbar p\,\partial_p
\end{eqnarray}
and it can generate the scaling transformation (\ref{st}) with the
choice $\gamma =i\,(\ln k)/\hbar$. Therefore if $u(q,p)$ is
expanded in power series it is easy to see that
\begin{eqnarray}
e^{\gamma \hat{L}_{M_3}}\,u(q,p)=u( k q,\,p/k ),
\end{eqnarray}
where we used the facts that
\begin{eqnarray}
e^{-i\hbar\gamma \, q\,\partial_q}\,q^n=(kq)^n,\quad
e^{i\hbar\gamma \, p\,\partial_p}\,p^n=(p/k)^n.
\end{eqnarray}
Consequently
\begin{eqnarray}\label{Laction}
F_L\star u(q,p)\star F_L^{-1}=u\left( aq+bp,cq+dp\right),
\end{eqnarray}
with
\begin{eqnarray}
a=(1+4\hbar^2\alpha\,\beta )k,\quad b=-2i\hbar\alpha/k,\quad
c=2i\hbar\beta k,\quad d=1/k.
\end{eqnarray}

If the same procedure is run for the decomposition
\begin{eqnarray}\label{fi}
F_I\,'(q,p)=e_\star^{iq^2/2\hbar}\,\star\,
e_\star^{ip^2/2\hbar}\,\star\,e_\star^{iq^2/2\hbar}
\end{eqnarray}
corresponding to the interchanging (\ref{inter}), it can be
concluded that
\begin{eqnarray}\label{interg}
e^{i(q^2+p^2)/\hbar}\star u(q,p)\star
[\,e^{i(q^2+p^2)/\hbar}\,]^{-1}=u(p,-q),
\end{eqnarray}
which is just as expected from the fact that the interchanging
(\ref{im}) is a special case of the linear CT (\ref{LMGF}). This
result relies on the uniqueness of the solution to (\ref{tech}),
i.e., the uniqueness of GF. But one should be aware that the
composition (\ref{fi}) is not a special case of (\ref{LCTGF}).
This situation is originated from the fact that because any finite
CT can be achieved by many different basic transformation steps,
the decomposition of the transformation is not unique.

Now this time consider the nonlinear CT
\begin{eqnarray}
Q(q,p)=q,\qquad P(q,p)=p+\nu q^2,
\end{eqnarray}
generated by
\begin{eqnarray}
F(q,p)=e^{-\frac{i\nu}{3\hbar}q^3},
\end{eqnarray}
where $\nu$ is a parameter. Thus the Lie operator is
\begin{eqnarray}
\hat{L}_M=q^3\star -\star
q^3=-\frac{1}{4}i\hbar^3\partial_p^{\;3}+3i\hbar\, q^2\partial_p .
\end{eqnarray}
Therefore $f(q,p)$ transforms as the following
\begin{eqnarray}\label{nonlin}
e^{-\frac{i\nu}{3\hbar}\hat{L}_M}\,f(q,p)&=&e^{-\frac{\nu
\hbar^2}{12}\partial_p^{\;3}+\nu\, q^2
\partial_p
}f(q,p)\nonumber\\
&=&e^{-\frac{\nu \hbar^2}{12}\partial_p^{\;3}}f(q,p+\nu q^2)\nonumber\\
&=&\left( 1-\frac{\nu \hbar^2}{12}\partial_p^{\;3}+\cdots
\right) f(q,p+\nu q^2)\nonumber\\
&=&f(q,p+\nu q^2)-\frac{\nu \hbar^2}{12}\partial_p^{\;3}f(q,p+\nu
q^2)+\cdots \;\;.
\end{eqnarray}
As a special case it may be remarkable to point out that one can
show easily that any gauge transformation $e^{\lambda f(q)}$ (
linear or nonlinear ) transforms $p^2$ and $p^{-1}$ as
$(p+i\hbar\lambda\partial_q f)^2$ and $(p+i\hbar\lambda\partial_q
f)^{-1}$ respectively. Note that the inverses mean the algebraic
inverses. If the interchange transformation is employed, the same
situation appears for $q^2$ and $q^{-1}$. Therefore this fact and
(\ref{Laction}) say that any gauge transformation can be applied
directly to the harmonic oscillator or Coulomb-type problems.

As a physical realization of the facts that have been stated so
far on the transform of functions, we will consider the
transformation of the $\star$-genvalue equation $H(q,p)\star
W(q,p)=E\,W(q,p)$ by a suitable example, where $W(q,p)$ is the
Wigner function of the system $H(q,p)$ and $E$ is the
correspondent eigenvalue \cite{ref:Fairlie}. A CT converts the
$\star$-genvalue equation into another $\star$-genvalue equation
$H'(q,p)\star W'(q,p)=E\,W'(q,p)$ where
\begin{eqnarray}
H'(q,p)=F(q,p)\star H(q,p)\star F^{-1}(q,p),\nonumber\\
W'(q,p)=F(q,p)\star W(q,p)\star F^{-1}(q,p).
\end{eqnarray}
Now consider the system with linear potential $H(q,p)=p^2+q$. The
Wigner function of the system satisfying the $\star$-genvalue
differential equation is the Airy function
\begin{eqnarray}
W(q,p)=Ai(\xi
)=\frac{1}{2\pi}\int_{-\infty}^{\infty}e^{i(t^3/3+\xi t)}dt,
\end{eqnarray}
where $\xi (q,p)=(2/\hbar )^{2/3}(q+p^2-E)$ \cite{ref:Curtright}.
The CT
\begin{eqnarray}
Q(q,p)=p-q^2,\quad P(q,p)=-q
\end{eqnarray}
that can be constructed by means of a two-step GF
\begin{eqnarray}
F(q,p)=e^{\frac{i\,q^3}{3\,\hbar}}\star
e^{\frac{i}{\hbar}(q^2+p^2)}
\end{eqnarray}
should convert the Hamilton and Wigner functions of the linear
potential system into that of the free particle system $H(q,p)=p$.
Indeed, the transformation of the Hamilton function is immediate
by (\ref{interg}) and (\ref{nonlin}). On the other hand, the first
step of the GF transforms the Wigner function as $Ai[\xi (p,-q)]$.
The second step acts as the following
\begin{eqnarray}
e^{\frac{i}{3\hbar}\hat{L}_M}\,Ai[\xi (p,-q)]&=&e^{\frac{
\hbar^2}{12}\partial_p^{\;3}-q^2\partial_p}Ai[\xi (p,-q)]\nonumber\\
&=&e^{\frac{
\hbar^2}{12}\partial_p^{\;3}}Ai[\xi (p-q^2,-q)]\nonumber\\
&=&(\hbar/2)^{2/3} \delta (p-E).
\end{eqnarray}
This is the free particle Wigner function.

\section{Intertwining}

Suppose that there exists a $c$-number phase space function
$L(q,p)$ making a link between two Hamilton functions in potential
form by means of the transformation
\begin{eqnarray}\label{intert}
L(q,p)\star H_0(q,p)\star L^{-1}(q,p)=H_1(q,p),
\end{eqnarray}
where $H_0=p^2+V_0(q)$ and $H_1=p^2+V_1(q)$. It is apparent that
(\ref{intert}) is equivalent to
\begin{eqnarray}\label{interwining}
L(q,p)\star H_0(q,p)=H_1(q,p)\star L(q,p),
\end{eqnarray}
and therefore $L(q,p)$ is an intertwining GF. Expansion of the
$\star$-product reduces (\ref{interwining}) to the differential
equation relating the two potentials;
\begin{eqnarray}\label{twopotentials}
V_1( q+i\hbar\partial_p/2 )\,L(q,p) = V_0( q-i\hbar\partial_p/2 )
\,L(q,p)+2i\hbar\, p\,\partial_q L(q,p).
\end{eqnarray}
The choice
\begin{eqnarray}\label{ansatz}
L(q,p)=p-i \varphi (q)
\end{eqnarray}
leads to the well-known consistency conditions
\begin{eqnarray}\
V_1(q)+V_0(q)=2\varphi^2(q),
\end{eqnarray}
\begin{eqnarray}\label{Riccati}
V_1(q)=V_0(q)+2\hbar \partial_q \varphi (q),
\end{eqnarray}
where $\varphi (q)$ is the solution to the Riccati equation
(\ref{Riccati}) which can be linearized by the Darboux
transformation $\varphi (q)=-\hbar\, \partial_q \phi (q)/\phi (q)$
to give the Schr\"{o}dinger equation with zero eigenvalue
\begin{eqnarray}
-\hbar^2 \partial_q^2 \phi (q)  +V_0(q)\phi (q)=0.
\end{eqnarray}

Now let us return to the ansatz (\ref{ansatz}). By (\ref{gauge}),
this is the gauge transformation
\begin{eqnarray}\label{p}
L(q,p)=e^{-\int \varphi (q)dq/\hbar}\star p\star e^{\int \varphi
(q)dq/\hbar}.
\end{eqnarray}
If we replace $p$ in (\ref{p}) by the decomposition
\begin{eqnarray}
p=e^{i(q^2+p^2)/\hbar}\star e^{\ln q}\star
[\,e^{i(q^2+p^2)/\hbar}\,]^{-1},
\end{eqnarray}
we conclude that $L(q,p)$ is a sequence of the fundamental
transformations and that intertwining is a CT. With the help of
(\ref{defnalt}), the explicit definition of the transformation
$L(q,p)$ is then given by the equations
\begin{eqnarray}
L(q,p)\star q \star L^{-1}(q,p)&=&q-i\hbar\,(p-i\varphi )^{-1},\\
L(q,p)\star p \star L^{-1}(q,p)&=&p+\hbar\,\partial_q \varphi\star
(p-i\varphi )^{-1}.
\end{eqnarray}

As a final remark note that (\ref{twopotentials}) can also be
used, besides the intertwining, to determine the GF for the given
any potential pair $V_0$ and $V_1$. For example for the pair
$V_0=q$ and $V_1=0$ that is a non-intertwining transformation from
the linear potential to the free particle, the solution is
\begin{eqnarray}
L(q,p)=e^{-2i\,(q\,p+4p^3/3)/\hbar}.
\end{eqnarray}
This fact is the most remarkable property of
(\ref{twopotentials}), that is it may relate any two systems
without considering their potentials are intertwined or not.

On the other hand, without regarding (\ref{twopotentials}), the
same transformation; i.e., $H_0=p^2+q \rightarrow H_1=p^2$ can be
obtained by the five-step sequence
\begin{eqnarray}\label{fivestep}
L(q,p)=e^{i(q^2+p^2)/\hbar}\star e^{\lambda [f(q)p+g(q)]}\star
[\,e^{i(q^2+p^2)/\hbar}\,]^{-1}\star e^{iq^3/3\hbar}\star
e^{i(q^2+p^2)/\hbar}
\end{eqnarray}
giving
\begin{eqnarray}
L(q,p)\star (p^2+q)\star L^{-1}(q,p)=p^2,
\end{eqnarray}
where
\begin{eqnarray}
f(q)=\frac{i}{\hbar\lambda}\left[2q+1+(1+8q)^{1/2}\right]
\end{eqnarray}
and
\begin{eqnarray}
g(q)=\frac{1}{2\lambda}\ln\left[\frac{1+(1+8q)^{1/2}}{1+8q}\right]
\end{eqnarray}
correspond to the transformation $Q=q^2$, $P=p/2q$. It seems at
first sight that the transformation (\ref{fivestep}) can be
converted easily into the $\star$-exponential form with the help
of (\ref{ptm}) and (\ref{fi}), but note that the fourth step
remains unclear since the determination of $f(q)$ for $A(q)=q^2$
in (\ref{A}) is not so easy.

\section{Summary and Conclusions}

In quantum mechanics a conjecture states that any quantum CT can
be generated as a sequence of three fundamental CTs. It is seen
that when the isomorphism (\ref{correspondence}) is used to write
the fundamental quantum CTs in $\star$-product formalism, gauge
and point transformations are immediate but the interchange is
not. But the system of differential equations (\ref{interchange})
allows us to get the generator of the interchange transformation
surprisingly in an ordinary exponential form. Parallel to this
result, it is shown that the others can also be obtained in the
ordinary form. The convertibility of (\ref{defn}) into a system of
differential equations allows us a powerful tool in determining
both the GF and the CT. Moreover, if point transformation is
defined by (\ref{ord-point}), the gauge transformation is
unnecessary and this reduces the number of independent
transformations to two.

It is also shown that the approach developed above offers results
compatible with the well-known behaviors of functions under linear
and nonlinear CTs. On the other hand, the intertwining method can
also be investigated within this framework. As an extra advantage,
(\ref{twopotentials}) offers a relation between any two systems
even though their potentials are not intertwined.

\section*{Acknowledgements}

This work was supported by T\"{U}B\.{I}TAK (Scientific and
Technical Research Council of Turkey) under contract 107T370.


\begin{thebibliography}{0}
\bibitem{ref:Weyl} H. Weyl,
  {\it Z. Phys.} {\bf 46}, 1 (1927).
\bibitem{ref:Wigner} E. Wigner,
  {\it Phys. Rev.} {\bf 40}, 749 (1932).
\bibitem{ref:Groenewold} H. Groenewold,
  {\it Physica} {\bf 12}, 405 (1946).
\bibitem{ref:Moyal} J. Moyal,
  {\it Proc. Camb. Phil. Soc.} {\bf 45}, 99 (1949).
\bibitem{ref:Schroek} F. E. Schroek, Jr.,
  {\it Quantum Mechanics on Phase Space}, (Kluwer, Dordrecht,
  1996).
\bibitem{ref:Dubin} D. A. Dubin, M. A. Hennings and T. D. Smith,
  {\it Mathematical Aspects of Weyl Quantization and Phase},
(World Scientific, Singapore, 2000).
\bibitem{ref:ZFC} C. K. Zachos, D. B. Fairlie and T. L. Curtright, Eds.,
  {\it Quantum Mechanics in Phase Space - An overview with selected papers}
(World Scientific, Singapore, 2005).
\bibitem{ref:Leaf} B. Leaf,
  {\it J. Math. Phys.} {\bf 9}, 65 (1968), {\it ibid} {\bf 9}, 769 (1968)
   and {\bf 10}, 1971 (1969).
\bibitem{ref:Leyvraz} F. Leyvraz and T. H. Seligman,
   {\it J. Math. Phys.} {\bf 30}, 2512 (1989).
\bibitem{ref:Deenen} J. Deenen,
  {\it J. Phys. A} {\bf 24}, 3851 (1991).
\bibitem{ref:Anderson} A. Anderson,
   {\it Ann. Phys}. {\bf 232}, 292 (1994), hep-th/9305054.
\bibitem{ref:Strat} R. L. Stratonovich,
  {\it Sov. Phys. JETP} {\bf 4}, 891 (1957).
\bibitem{ref:Brif} C. Brif and A. Mann,
  {\it Phys. Rev. A} {\bf 59}, 971 (1999).
\bibitem{ref:TH} T. Hakio\u{g}lu, A. Te\u{g}men and B. Demircio\u{g}lu,
  {\it Phys. Lett. A} {\bf 360}, 501 (2007), quant-ph/0605236.
\bibitem{ref:Imre} K. Imre {\it et al.},
  {\it J. Math. Phys}. {\bf 8}, 1097 (1967).
\bibitem{ref:Fronsdal} C. Fronsdal,
  {\it Rep. Math. Phys}. {\bf 15}, 111 (1978).
\bibitem{ref:Bayen} F. Bayen {\it et al.},
  {\it Ann. Phys.} {\bf 111} 61, (1978), {\it ibid} {\bf 111} 111,
  (1978).
\bibitem{ref:THmain} T. Hakio\u{g}lu,
  Extented covariance under nonlinear canonical
  transformations in Weyl quantization, quant-ph/0011076.
\bibitem{ref:JOSAA} T. Hakio\u{g}lu,
{\it J. Opt. Soc. Am. A} {\bf 17} 2411, (2000).
\bibitem{ref:Dragt} A. J. Dragt and S. Habib,
How Wigner functions transform under symplectic maps,
quant-ph/9806056.
\bibitem{ref:Curtright} T. Curtright, D. Fairlie and C. Zachos,
{\it Phys. Rev. D} {\bf 58} 025002 (1998), hep-th/9711183.
\bibitem{ref:Zachos} C. Zachos,
  {\it Int. J. Mod. Phys. A} {\bf 17} 297, (2002),
  hep-th/0110114.
\bibitem{ref:ADTH} T. Hakio\u{g}lu and A. Dragt,
{\it J. Phys. A} {\bf 34} 6603, (2001), quant-ph/0108081.
\bibitem{ref:Fairlie} D. Fairlie,
{\it Cambridge Philos. Soc.} {\bf 60} 581, (1964).
\end{thebibliography}
\end{document}